\documentclass[showpacs,pra,twocolumn]{revtex4}
\usepackage{epsfig,amsmath}
\usepackage{subfigure}
\usepackage{graphicx}
\usepackage{amsthm}
\usepackage{amssymb}
\usepackage{bm}
\usepackage{hyperref}
\usepackage{color}
\newcommand{\beq}{\begin{equation}}
\newcommand{\eeq}{\end{equation}}
\newcommand{\beqa}{\begin{eqnarray}}
\newcommand{\eeqa}{\end{eqnarray}}

\begin{document}

\title{Quantum energy transfer between nonlinearly-coupled bosonic bath and a fermionic chain: an exactly solvable model}
\author{Zhao-Ming Wang$^{1,2}$}
\author{Da-Wei Luo$^{4}$}
\author{Baowen Li$^{5}$}
\author{Lian-Ao Wu$^{2,3}$}
\email{Corresponding author: lianao.wu@ehu.es}
\affiliation{$^{1}$ Department of Physics, Ocean University of China, Qingdao, 266100,
China}
\affiliation{$^{2}$ Department of Theoretical Physics and History of Science, The Basque
Country University(EHU/UPV), 48008, Spain}
\affiliation{$^{3}$ IKERBASQUE, Basque Foundation for Science, 48011 Bilbao, Spain}
\affiliation{$^{4}$ Center for Quantum Science and Engineering, and Department of
Physics, Stevens Institute of Technology, Hoboken, New Jersey 07030, USA}
\affiliation{$^{5}$ Department of Mechanical Engineering, University of Colorado,
Boulder, CO 80309}
\date{\today }

\begin{abstract}
The evolution of a quantum system towards thermal equilibrium is usually studied by approximate methods, which have their limits of validity and should be checked against analytically solvable models. In this paper, we propose an analytically solvable model to investigate the energy transfer between a bosonic bath and a fermionic chain which are nonlinearly-coupled to each other. The bosonic bath consists of an infinite collection of non-interacting bosonic modes, while the fermionic chain is represented by a chain of interacting fermions with nearest-neighbor interactions. We compare behaviors of the temperature-dependent energy current $J_{T}$ and temperature-independent energy current $J_{TI}$ for different bath configurations. With respect to the bath spectrum, $J_{T}$ decays exponentially for Lorentz-Drude type bath, which is the same as the conventional approximations. On the other hand, the decay rate is $1/t^{3}$ for Ohmic type and $1/t$ for white noise, which doesn't have conventional counterparts. For the temperature-independent current $J_{TI}$, the decay rate is divergent for the Lorentz-Drude type bath, $1/t^{4}$ for the Ohmic bath, and $1/t$ for the white noise. When further considering the dynamics of the fermionic chain, the current will be modulated based on the envelope from the bath. As an example, for a bosonic bath with Ohmic spectrum, when the fermionic chain is uniformly-coupled, we have $J_{T}\propto1/t^{6}$ and $J_{TI}\propto1/t^{3}$. Remarkably, for perfect state transfer (PST) couplings, there always exists an oscillating quantum energy current $J_{TI}$. Moreover, it is interesting that $J_{T}$ is proportional to $(N-1)^{1/2}$ at certain times for PST couplings under Lorentz-Drude or Ohmic bath.
\end{abstract}

\maketitle


\section{Introduction}

Dissipative phenomena in open systems~\cite{Legget,Caldrira} can give rise to a variety of interesting physical scenarios and have been under extensive study across many different fields such as quantum optics, many-body physics, and quantum information sciences. Open systems are notoriously difficult to deal with exactly due to the complexity of the quantum reservoir, whose Hilbert space can be prohibitively large. Such systems are usually tackled with a system-plus-reservoir approach where one treats the composite system as a whole, and later traces out the reservoir degrees of freedom to study the reduced dynamical behaviors of the system under consideration. One widely-used way to model the quantum reservoir is to treat it as non-interacting harmonic oscillators~\cite{Hedegard,Neto}. The dynamics of two Brownian particles in a common reservoir have been studied~\cite{Duarte} as well as its thermal equilibrium properties~\cite{Valente}. Using a quantum Langevin description, a
system-reservoir model is proposed~\cite{Bhattacharya} and a quantum current
is observed which is dependent on various parameters of external noise.
Spin-boson model also provides a clear physical picture for exploring
quantum dissipation effects. This model includes an impurity two-level
system (TLS) coupled to a thermal reservoir and displays a
rich phase diagram in the equilibrium regime~\cite{Legget, Hur}. A
generalized non-equilibrium polaron-transformed Redfield equation with an
auxiliary counting field was developed recently to study the full counting
statistics of quantum heat transfer in a driven non-equilibrium spin-boson
model~\cite{Chen17}. For a subsystem which constitutes of two interacting spins, this situation
effectively corresponds to a subsystem unharmonically coupled to a bosonic
bath, allowing to introduce nonlinear effects~\cite{Vierheilig}. Thermal rectification within a spin-boson nanojunction model is analyzed and analytic solutions are obtained for a separable model and a nonseparable model \cite{Divra}. Exact
dynamics of interacting TLS immersed in separate thermal reservoirs or within a common bath has also been studied~\cite{Wu2013}. For the device design, such as in molecular devices, people often need to consider the scaling of heat current with system size and time in order to prevent the devices from disintegrating~\cite{Schulze,Pop}, because excess heat build-up during operation may cause device disintegrating.

However, most theoretical investigations of how a quantum system reaches thermal
equilibrium use the approximation methods, such as quantum master
equations~\cite{Wu09,Cao18,Ramezani}, Born-Oppenheimer methods~\cite{Wu11}, etc. Typically, such methods only provide
numerical results, hindering a direct picture of the microscopic processes
involved. Responding to this challenge, we have recently developed an
analytic method for describing the energy transfer in a hybrid quantum system.
The hybrid quantum system consists of a bosonic bath and a fermionic chain,
which are nonlinearly coupled by using a dressing transformation. Physically this prototype quantum model could be realized in different system. For example, the bosonic bath and the fermionic chain correspond to harmonic solids and metal (or spin) respectively. In Ref \cite{Lianao}, thermal rectification appears in two different reservoirs connected by molecular vibration. Our results show that the energy current can disappear at some times for this nonlinearly coupled hybrid bath, whereas for two linearly coupled bosonic bath, a steady current will be obtained~\cite{Bonetto}.

\section{Model}


Consider two baths ($H_{LB}$ and $H_{RB}$) connected by a central system $H_{S}$. The central system $S$ consists of two interacting fermions. The left bath $LB$ is modeled as a collection of non-interacting bosonic modes~\cite{Hedegard,Neto}, maintained at a fixed temperature $T=\beta ^{-1}$, with $k_{B}=1$. The right bath $RB$ is modeled as a one-dimensional fermionic chain with nearest-neighbor interactions. The total Hamiltonian is given by
\begin{equation}
H=H_{S}+H_{LB}+H_{RB}+V_{L}+V_{R},
\end{equation}
where $V_{L}$ ($V_{R}$) is the interaction between the left (right) bath and the central system. Treating the central system and the fermionic right bath as a whole, we denote $H_{ch}=H_{S}+H_{RB}+V_{R}$ and rearranged the indices so that the interacting fermions in central system is labeled $1$ and $2$ and the sites in the fermionic chain are labeled $3$ through $N$,
\begin{equation}
H_{ch}=H_{S}+H_{RB}+V_{R}=-\sum\limits_{i=1}^{N-1}\tau _{i}(c_{i}^{\dag }c_{i+1}+c_{i+1}^{\dag
}c_{i}),
\end{equation}
where $H_{S}=-\tau _{1}(c_{1}^{\dag }c_{2}+c_{2}^{\dag
}c_{1})$, $H_{RB}=-\sum\limits_{i=3}^{N-1}\tau _{i}(c_{i}^{\dag }c_{i+1}+c_{i+1}^{\dag
}c_{i})$, $V_{R}=-\tau _{2}(c_{2}^{\dag }c_{3}+c_{3}^{\dag
}c_{2})$, $c_{i}^{\dag }$ is standard fermionic creation operator of electron spin, and $\tau_{i}$ is the coupling constant between the nearest neighbor sites. Additionally, we take $\tau _{i}>0$ which corresponds to ferromagnetic couplings throughout.  $H_{ch}$ describes a chain of interacting spin-less fermions, which can be mapped to a one-dimensional XY chain under Jordan-Wigner transformation. This is true for an open-ended chain or a periodic chain \cite{AOP,JPA2004,PRA2004}. Here we consider an open-ended chain. For simplicity, the total Hamiltonian $H$ can be written as  (see Fig. 1)
\begin{equation}
H=H_{ch}+H_{B}+V,
\label{equat1}
\end{equation}
where $H_{B}=H_{LB}=\sum\nolimits_{\alpha }\omega _{\alpha }b_{\alpha }^{\dag }b_{\alpha }$ is the bosonic bath's Hamiltonian (or phonons if zero point energy is added). $b_{i}^{\dag }$ is the bosonic creation operator, and $\omega _{\alpha }$ is the
frequency of the $\alpha$-th mode. Note that the total number of excitations $M=\sum_{i}c_{i}^{\dagger}c_{i}$ in the chain remains constant, and thus the $z$ component of the total spin is a conserved quantity. We can discuss problems in a fixed subspace for certain excitations. For simplicity, we only consider $M=1$ case.  We will study the
energy transfer between the bath and the chain next. The energy current will be zero after some time $t$ for a finite length chain, at which point thermal equilibrium is reached.

\begin{figure}[]
\centerline{\includegraphics[width=1.0\columnwidth]{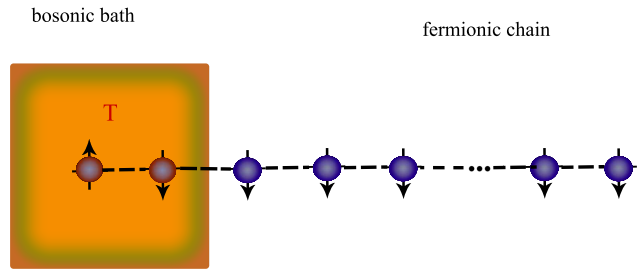}}
\caption{(Color on line) A schematic representation of the model. The bosonic bath and the fermionic chain are connected by two fermions. The fermionic chain can be mapped to a one-dimensional spin chain and the bosonic bath is at a finite temperature $T$ while the fermionic chain is at zero temperature.}
\label{fig:1}
\end{figure}

A specific interaction $V=V_{L}$ can be introduced by the so-called
dressing transformation $W^{\dag }(H_{ch}+H_{B})W$ with a displacement
operator $W=\exp [\sum\nolimits_{\alpha }(\Gamma _{\alpha }b_{\alpha }^{\dag
}-\Gamma _{\alpha }^{\ast }b_{\alpha })n_{1}]$, where $n_{i}=c_{i}^{\dag }c_{i}$ is the number
operator. Clearly, it is possible to introduce different types of interactions using different types of dressing
transformations~\cite{Wu09}. the exact form of the interaction operator is now given by
\begin{eqnarray}
V &=&\sum\nolimits_{\alpha }\omega _{\alpha }(\Gamma _{\alpha }^{\ast
}b_{\alpha }+\Gamma _{\alpha }b_{\alpha }^{\dag }+\left\vert \Gamma _{\alpha
}\right\vert ^{2})n_{1}  \notag \\
&&+\tau _{1}(c_{1}^{\dag }c_{2}e^{\Gamma _{\alpha }^{\ast }b_{\alpha
}-\Gamma _{\alpha }b_{\alpha }^{\dag }}-1+h.c.).
\end{eqnarray}
$\Gamma _{\alpha }$
parameterizes the system-bath coupling strength, and is assumed to be a complex number.


\section{The calculation of energy expectation value and energy current}

In this section, we will give a general analysis of the energy current between the bosonic bath and the fermionic chain. The energy current operator can be defined as~\cite{Wu09}
\begin{equation}
J=i[V,H_{B}].
\end{equation}
The expectation value of the energy current is given by
\begin{equation}
J(t)=Tr[\rho _{0}\widehat{J}],
\end{equation}
where $\rho _{0}=\rho_{B}\otimes \rho _{ch}$ is the initial density operator of the whole system and $\widehat{J}%
=e^{iHt}\widehat{J}(0)e^{-iHt}$. Equivalently, the energy current may be rewritten as
\begin{equation}
J(t)=\frac{\partial \left\langle H_{B}(t)\right\rangle }{\partial t}
\end{equation}%
where $\left\langle H_{B}(t)\right\rangle =Tr(\rho _{0}\widehat{H_{B}}(t))$ is the energy expectation value
of the bath, where $\widehat{H_{B}}$=$e^{iHt}\widehat{H}_{B}(0)e^{-iHt}$. Note that a positive value represents an energy current from bath to the chain and vice versa. The bosonic bath at temperature $T$ is modeled as a canonical ensemble with distribution $\rho _{B}=\exp
(-\beta H_{B})/Tr[\exp (-\beta H_{B})]$.

It can be readily shown that
\begin{equation}
U_{0}^{\dag }b_{\alpha }^{\dag }U_{0}=b_{\alpha }^{\dag }e^{i\omega _{\alpha
}t},
\end{equation}
and
\begin{equation}
U_{0}^{\dag }c_{1}^{\dag }U_{0}=c_{1}^{\dag
}(t)=\sum\limits_{l}f_{1,l}^{\ast }c_{l}^{\dag },
\end{equation}%
where $U_{0}=e^{-i(H_{B}+H_{ch})t}$ and $f_{1,l}$ is the transition amplitude
of an excitation (the $\left\vert 1\right\rangle $ state) from site $1$ to
site $l$ in the chain. We restrict the fermion chain to have only one excitation. Denoting $\left\vert \overline{l}\right\rangle$ as the state where the one fermion excitation is at site $l$, we write the initial state of the fermion chain as $\left\vert
\Psi (0)\right\rangle =\frac{1}{\sqrt{a}}\sum\limits_{l=1}^{a}\left\vert
\overline{l}\right\rangle$, where the 1 excitation is restricted to the first $a$ sites. Then, $\left\langle
H_{B}(t)\right\rangle $ can be expressed as (neglecting the time-independent
part)

\begin{equation}
\left\langle H_{B}(t)\right\rangle =\left\langle H_{B}(t)\right\rangle
_{T}+\left\langle H_{B}(t)\right\rangle _{TI},
\end{equation}%
where $\left\langle H_{B}(t)\right\rangle _{T}$ is the temperature dependent
part

\begin{eqnarray}
&&\left\langle H_{B}(t)\right\rangle _{T}  \notag \\
&=&\frac{1}{a}\sum\nolimits_{\alpha }\omega _{\alpha }\left\vert \Gamma
_{\alpha }\right\vert ^{2}\left\langle D(\Gamma )\right\rangle _{eq}\{2\coth
\frac{\beta \omega _{\alpha }}{2}\sin \omega _{\alpha }t\mathtt{Im}[F(t)]
\notag \\
&&+2(1-\cos \omega _{\alpha }t)\mathtt{Re}[F(t)]\},
\end{eqnarray}

and $\left\langle H_{B}(t)\right\rangle _{TI}$ is the temperature
independent part

\begin{eqnarray}
&&\left\langle H_{B}(t)\right\rangle _{TI}  \notag \\
&=&\frac{1}{a}\sum\nolimits_{\alpha }\omega _{\alpha }\left\vert \Gamma
_{\alpha }\right\vert ^{2}[(1-2\cos \omega _{\alpha }t)\left\vert
f_{11}\right\vert ^{2}+G(t)],
\end{eqnarray}
where $\left\langle D(\Gamma )\right\rangle _{eq}=\exp (-\frac{1}{2}%
\sum\nolimits_{\alpha }\left\vert \Gamma _{\alpha }\right\vert ^{2}\coth
\frac{\beta \omega _{\alpha }}{2})$ is the expectation value of the
displacement operator in the thermal equilibrium state, and $F(t),G(t)$ indicate
the dynamics of the chain and depend on the initial state of the chain. The explicit definition of $F(t)$ and $G(t)$ are given for three different initial states in the following sections. Next we will discuss the behavior of the energy current for different initial states, different spectra for the bosonic bath and different coupling configurations in the fermionic chain.


\section{Energy current for different initial states of the chain}

Now we discuss the energy current for some different initial states. The energy current in the contact can also be expressed as two parts

\begin{equation}
J=J_{T}+J_{TI},
\end{equation}%
where
\begin{eqnarray}
J_{T} &=&\frac{2}{a}\sum\nolimits_{\alpha }\omega _{\alpha }\left\vert
\Gamma _{\alpha }\right\vert ^{2}\left\langle D(\Gamma _{\alpha
})\right\rangle _{eq}  \notag  \label{eq:14} \\
&&\{\coth \frac{\beta \omega _{\alpha }}{2}[\omega _{\alpha }\cos \omega
_{\alpha }t\mathtt{Im}(F(t))  \notag \\
&&+\sin \omega _{\alpha }td\mathtt{Im}(F(t))/dt]+\omega _{\alpha }\sin
\omega _{\alpha }t\mathtt{Re}(F(t))  \notag \\
&&+(1-\cos \omega _{\alpha }t)d\mathtt{Re}(F(t))/dt\},
\end{eqnarray}
is the temperature-dependent current, and

\begin{eqnarray}
J_{TI} &=&\frac{1}{a}\sum\nolimits_{\alpha }\omega _{\alpha }\left\vert
\Gamma _{\alpha }\right\vert ^{2}[2\omega _{\alpha }\sin \omega _{\alpha
}t\left\vert f_{11}\right\vert ^{2}+   \\  \notag
&&(1-\cos \omega _{\alpha }t)d(\left\vert f_{11}\right\vert
^{2})/dt+dG(t)/dt],
\label{eq:16}
\end{eqnarray}
is the temperature-independent current.

(i) $\left\vert \Psi (0)\right\rangle =\left\vert \overline{1}\right\rangle
$. In this case, $a=1,F(t)=0,G(t)=0$, and $J_{T}=0$. Therefore, the current is
independent of the temperature and represents a pure quantum current. We also note that if the initial state
is prepared as $\alpha \left\vert \overline{0}\right\rangle +\beta
\left\vert \overline{1}\right\rangle $, i.e., the first site is an arbitrary pure
state $\alpha \left\vert 0\right\rangle +\beta \left\vert 1\right\rangle $
and all other sites at state $\left\vert 0\right\rangle $, the state on site
one can not be transferred to other sites under the chain dynamics.

(ii) $\left\vert \Psi (0)\right\rangle =\frac{1}{\sqrt{N}}(\left\vert
\overline{1}\right\rangle +\left\vert \overline{2}\right\rangle
+...+\left\vert \overline{N}\right\rangle )$, now $a=N,F(t)=f_{1,1}^{\ast
}\sum\limits_{l=2}^{N}f_{1,l},G(t)=\sum\limits_{l,m=2}f_{1,l}^{\ast }f_{1,m}$%
.

(iii) $\left\vert \Psi (0)\right\rangle =\frac{1}{\sqrt{2}}(\left\vert
\overline{1}\right\rangle +\left\vert \overline{2}\right\rangle )$, $%
a=2,F(t)=f_{1,1}^{\ast }f_{12},G(t)=\left\vert f_{1,2}\right\vert ^{2}$.

Note that the analytical expression of the energy current above is obtained without approximations. In the next section, we will
discuss case (iii) as an illustrative example.

\section{Effects of the spectrum distribution for the bosonic bath}

When the chain-bath couplings are weak $(\Gamma _{\alpha }/\tau _{\alpha
}\rightarrow 0)$, which corresponds to the Markovian limit, the displacement
operator expectation value in the thermal equilibrium can be approximated its
the first order term $\exp (-\frac{1}{2}\sum\nolimits_{\alpha }\left\vert
\Gamma _{\alpha }\right\vert ^{2}\coth \frac{\beta \omega _{\alpha }}{2}%
)\approx 1$. Additionally, in the high temperature $T$, or low frequency limit $\omega
_{\alpha }\rightarrow 0$, the hyperbolic cotangent $\coth \frac{\omega _{\alpha }%
}{2T}\rightarrow \infty $, so the dominant term will be the first two terms in
Eq.~(\ref{eq:14}). Using the Taylor expansion for the hyperbolic
cotangent $\coth x=\sum\limits_{n=1}^{\infty }\frac{2^{2n}B_{2n}x^{2n-1}}{%
(2n)!}$ (where $0<\left\vert x\right\vert <\pi$ and $B_{n}$ is the $n$th Bernoulli number)
and taking the first order for $x\rightarrow 0$, the current in Eq.~(%
\ref{eq:14}) can be further reduced to

\begin{eqnarray}
J_{T} &=&2T\sum\nolimits_{\alpha }\left\vert \Gamma _{\alpha }\right\vert
^{2}[\sin \omega _{\alpha }t\frac{d\mathtt{Im}(F(t))}{dt}  \notag \\
&&+\omega _{\alpha }\cos \omega _{\alpha }t\mathtt{Im}(F(t))].
\end{eqnarray}%

Denoting $\Gamma$ as an overall coupling strength factor, the energy current $J_{T}, J_{TI}$ for different spectrum distribution
of the bath can then be written in this simplified form.

(i) Lorentz-Drude type bath~\cite{Wangjiansheng2013}, whose the spectrum density $\rho
(\omega )=\frac{\omega }{\omega _{d}^{2}+\omega ^{2}}$. The temperature-dependent energy current is given by
\begin{equation}
J_{T}=\pi T\left\vert \Gamma \right\vert ^{2}e^{-\omega _{d}t}[\frac{d%
\mathtt{Im}(F(t))}{dt}-\omega _{d}\mathtt{Im}(F(t))].  \label{eq:21}
\end{equation}%
In this case, the current decays exponentially with time $t$, modulated by $%
F(t)$ which comes from the chain's dynamics and finally it reaches
zero. Then the thermal equilibrium state is obtained. The envelope is an
exponentially decreasing line. For $J_{TI}$, it is divergent in Lorentz-Drude
bath.

(ii) Ohmic bath, with a spectrum density $\rho (\omega )=\frac{\pi }{2}\omega
e^{-\omega /\omega _{c}}$, where $\omega _{c}$ is the cut-off frequency.

For a long time limit $(\omega_{c}t)\gg 1$,

\begin{equation}
J_{T}\approx \frac{2\pi T\left\vert \Gamma \right\vert ^{2}}{\omega _{c}t^{4}%
}[t\frac{d\mathtt{Im}(F(t))}{dt}-3\mathtt{Im}(F(t))],  \label{eq:24}
\end{equation}%
\begin{eqnarray}
J_{TI} &\approx &\frac{\pi \left\vert \Gamma \right\vert ^{2}}{2}[\frac{3}{%
t^{4}}\left\vert f_{1,1}(t)\right\vert ^{2}+\frac{6}{\omega _{c}t^{4}}%
d(\left\vert f_{1,1}(t)\right\vert ^{2})/dt  \notag \\
&&+\omega _{c}^{3}d(\left\vert f_{1,1}(t)\right\vert ^{2}+\left\vert
f_{1,2}(t)\right\vert ^{2})/dt].
\label {eq19}
\end{eqnarray}%
Therefore, for $J_{T}$ the envelope becomes $1/t^{3}$ while for $J_{TI}$ it is $%
1/t^{4}$ for Ohmic baths. The quantum current decreased more quickly
than classical current for the Ohmic bath. Note that for $J_{TI}$, in the long
time limit, the current $J_{TI}$ will only depend on the chain's dynmics.

(iii) A ``white-noise'' spectrum where the frequency distributes uniformly with a cut-off
frequency
\begin{equation*}
\rho (\omega )=
\begin{cases}
1 & (0<\omega \leq \Omega ) \\
0 & (\text{otherwise} )%
\end{cases},
\end{equation*}%
then the energy current in the long time limit ($t \gg 1$) is given by
\begin{eqnarray}
J_{T} &\approx &2T\left\vert \Gamma \right\vert ^{2}\{[\frac{\Omega }{t}\sin
\Omega t]\mathtt{Im}(F(t))  \notag \\
&&+\frac{(1-\cos \Omega t)}{t}\frac{d\mathtt{Im}F(t)}{dt}\},
\end{eqnarray}

\begin{eqnarray}
J_{TI} &\approx &\left\vert \Gamma \right\vert ^{2}\{[\frac{\Omega ^{2}\sin
\Omega t}{t}]\left\vert f_{1,1}(t)\right\vert ^{2}  \notag \\
&&-\frac{\Omega \sin \Omega t}{t}d(\left\vert f_{1,1}(t)\right\vert ^{2})/dt
\notag \\
&&+\frac{\Omega ^{2}}{4}d(\left\vert f_{1,1}(t)\right\vert ^{2}+\left\vert
f_{1,2}(t)\right\vert ^{2})/dt\}.
\end{eqnarray}

The envelope for the currents is $1/t$. Note that for $J_{TI}(t)$ the current
depends on the chain's dynamics only in the long time limit.


\section{Effects of the fermionic chain configuration}
Now we reveal how different configurations of the fermionic chain can affect the energy current. The transition
amplitude $f_{m,m'}$ depends on the types of couplings in the chain. First we discuss the perfect state transfer (PST)
couplings $\tau _{k}=2\tau \sqrt{k(N-k)/N^{2}}$. The transition
amplitude reads
\begin{equation}
f_{m,m^{^{\prime }}}(t)=\exp [i\frac{\pi }{2}(m-m^{^{\prime
}})]d_{m^{^{\prime }},m}^{l}(2\tau t),
\end{equation}
where $d_{m^{^{\prime }},m}^{l}(2\tau t)$ is the Wigner \emph{d} matrix~\cite%
{Biedenharn}. The indices of the site number of a 1-dimensional chain can
be mapped onto the magnetic quantum numbers $m$ of the total angular
momentum $l$, such that $l=\frac{N-1}{2},m=-\frac{N-1}{2}+k-1$, where $k=1,2,...,N$ ~\cite{prl}. Using this relation, we obtain $f_{1,1}(t)=[\cos
\tau t]^{N-1},f_{1,2}(t)=i\sqrt{N-1}\sin \tau t[\cos \tau
t]^{N-2},f_{1,N}(t)=(-1)^{N}\exp [i\frac{\pi }{2}(N-1)][\sin \tau t]^{N-1}$. Note that the transition amplitude $f_{1,n}(t)$ $(n=1,2,N)$ are
periodic functions. When transferring a quantum state from one end to
the other end of the chain, the transmission fidelity is a function of
the transition amplitude and it can periodically reach 1. This is the so-called perfect state transfer (PST) \cite{prl}.

For the Lorentz-Drude spectrum, from Eq.~(\ref{eq:21}), the current is given by
\begin{eqnarray}
J_{T} &=&\sqrt{N-1}\pi T\left\vert \Gamma \right\vert ^{2}e^{-\omega _{d}t}
\notag \\
&&\{\tau \lbrack \cos (\tau t)]^{2(N-1)} \\
&&-\tau (2N-3)\sin ^{2}\tau t[\cos (\tau t)]^{2(N-2)}]  \notag \\
&&-\omega _{d}\sin (\tau t)[\cos (\tau t)^{2N-3}]\},  \notag
\end{eqnarray}

It is interesting to note that when $t=2n\pi /\tau $, $n=1,2,$..., $\cos
(\tau t)=1,\sin (\tau t)=0$, so we have

\begin{equation}
J_{T}=\sqrt{N-1}\tau \pi T\left\vert \Gamma \right\vert ^{2}e^{-2\omega
_{d}n\pi /\tau }.
\end{equation}
i.e. $J_{T}\varpropto (N-1)^{\frac{1}{2}}$. Note that $%
[\cos (\tau t)]^{N}$ becomes a $\delta $ function for big $N$. The current
will appear suddenly when $t=2n\pi /\tau $, and disappear at other $t$, this
behavior will be modulated by the exponential decreasing which comes from
the thermal bath. A larger bath size will absorb more energy for
the PST couplings.

For Ohmic spectrum at $t=2n\pi /\tau $, from Eq.~(\ref{eq:24}), $%
J_{T}\varpropto (N-1)^{\frac{1}{2}}$. For $J_{TI}$ at long time limit, it is
proportional to
\begin{widetext}
\[
\frac{\tau }{4}(N-1)[\cos 2\tau t(\cos \tau
t)^{2N-6}-(2N-6)\sin \tau t\sin 2\tau t(\cos \tau t)^{2N-8}]-2\tau (N-1)\sin
\tau t[\cos \tau t]^{2(N-1)-1}.
\]
\end{widetext}
Then there exists an oscillating quantum current $J_{TI}$ for PST couplings with an Ohmic bath, whose time average vanishes. This could happen in pure dephasing models where system bath coupling commutes with system Hamiltonian or if there are some symmetries that lead to multiple steady states. However our model does not satisfy the above two cases. Thus it might be inappropriate to consider the long time $t$ limit here. The underlying physics could be a quantum effect similar to persistent alternating electric current or the superfluid current in Ref. \cite{PRA20}.

For uniform couplings $\tau _{i}=\tau /2$, the transition amplitudes $f_{j,l}
$ from site $j$ to $l$ is,
\begin{equation}
f_{j,l}=\frac{2}{N+1}\sum\limits_{m=1}^{N}\sin (q_{m}j)\sin
(q_{m}l)e^{iE_{m}t},
\end{equation}
where $q_{m}=\pi m/(N+1),E_{m}=-\tau \cos q_{m}$.

Note that the transition probability $f_{1,l}(t)$ is real when $l$
is odd, and imaginary when $l$ is even. That is a typical odd-even effect and it
is a universal properties for finite systems~\cite{Wu97}. The more evident
effects will be displayed for smaller $N$. Clearly when $N=2$, $\mathtt{%
Re}(f_{1,1}^{\ast }f_{1,2})=0.$

When $N$ is infinite, the transition amplitude can be calculated as $f_{1,1}=%
\frac{1}{2}[J_{0}(\tau t)+J_{2}(\tau t)]$, where $J_{n}(t)$ is the Bessel
function of the first kind. when $l>1$ and $n=0,1,2,...$

\begin{equation}
f_{1,l}=
\begin{cases}
\frac{1}{\tau t}lJ_{l}(\tau t), & l=4n+1 \\
-\frac{i}{\tau t}lJ_{l}(\tau t), & l=4n+2 \\
-\frac{1}{\tau t}lJ_{l}(\tau t), & l=4n+3 \\
\frac{i}{\tau t}lJ_{l}(\tau t), & l=4n+4%
\end{cases}
\end{equation}

From the expression of $f_{1,1}$ and $f_{1,2}$, we can see that for both PST couplings and
uniform couplings, $f_{1,1}$is real and $f_{1,2}$ is imaginary. Then Re$%
[F(t)]\equiv 0(F(t)=f_{1,1}^{\ast }f_{1,2})$ in Eq.~(\ref{eq:14}). Thus even
if we do not consider the high temperature or low frequency, the last two
terms can be neglected for PST or uniform couplings. Using Ohmic spectrum for the bosonic bath
as an example,

\begin{equation}
J_{T}\approx \frac{2\pi T\left\vert \Gamma \right\vert ^{2}}{\omega _{c}}\{%
\frac{2J_{1}J_{3}-J_{2}[J_{0}-J_{2}]}{4\tau ^{2}t^{5}}+\frac{6J_{1}J_{2}}{%
\tau ^{2}t^{6}}\},
\end{equation}

\begin{eqnarray}
J_{TI} &\approx &\frac{\pi \left\vert \Gamma \right\vert ^{2}}{2}[\frac{3}{%
t^{4}}\left\vert f_{1,1}(t)\right\vert ^{2}+\frac{6}{\omega _{c}t^{4}}%
d(\left\vert f_{1,1}(t)\right\vert ^{2})/dt  \notag \\
&&+\omega _{c}^{3}d(\left\vert f_{1,1}(t)\right\vert ^{2}+\left\vert
f_{1,2}(t)\right\vert ^{2})/dt].
\end{eqnarray}

When $t\rightarrow \infty ,J_{n}(t)\approx \sqrt{\frac{2}{\pi t}}\cos (t-%
\frac{n\pi }{2}-\frac{\pi }{4})$, so $J_{T}\propto 1/t^{6}=(1/t^{3})^{2}$, where the
the bath spectrum and chain's dynamics both contribute a factor of $1/t^3$.
On the other hand, $J_{TI}\propto 1/t^{3}$ only, which corresponds to the dissipation in the
uniform chain. Here we would like to point out that when $N\rightarrow \infty$ and $t\rightarrow \infty$, it is shown that the asymptotic behaviours of $J_{T}$ might be unclear due to complexities of asymptotic processes, for instance the order of taking limits $N\rightarrow \infty$ and $t\rightarrow \infty$. A simple analytical asymptotic expression of the PST
couplings sheds light on the issue, $J_T\propto \mid\Gamma\mid^{2} \sqrt{N}/t^{3}$ is allowed to be a finite value if $N\rightarrow \infty$ and $t^3\rightarrow \infty$ share the same asymptotic speeds. The underlying physics for the possibility of the existence of this finite current and the relation between the asymptotic behaviour and the Markovian assumption need further research. The ratio of $J_{T}(t)/J_{TI}(t)\approx \frac{T\tan (\tau t-\frac{\pi }{4})}{2\tau \omega _{c}^{4}t^{3}}$ is proportional to
temperature $T$, couplings intensity $1/\tau $, cut-off frequency $1/\omega
_{c}^{4}$, and time $1/t^{3}$ modulated by a $tan$ function.


\section{Conclusions}

We analytically calculate the energy current between a bosonic bath and a fermionic chain. For our system, only an initial
entanglement state in the chain can induce a temperature dependent energy current.
The energy current $J(t)$ depends on both the bosonic bath spectrum and the coupling
mechanisms within the fermion chain. With respect to the effects of the bath spectrum, $J_{T}$ will
decrease to zero, with an exponential decay for Lorentz-Drude type which is in
accordance with the conventional Markovian approximation. On the other hand, it is proportional to $1/t^{3}$ for Ohmic spectrum, and $1/t$ for white noise. For $J_{TI}$, the effect
of the bath spectrum becomes divergent, $1/t^{4}$ and $1/t$, respectively.
When different coupling configurations in the fermion chain are introduced, the envelope of the energy current will be modulated. For PST couplings, the
oscillation is governed by a periodical function and for
uniform couplings it is governed by the Bessel function of the first kind. Additionally, for PST couplings, with a Lorentz-Drude or Ohmic bosonic bath, $J_{T}$ is found to be proportional to ($N-1)^{1/2}$ at certain times. This behavior can be interpreted as larger baths have the capacity to absorb more energy. To the best of our knowledge, our work is the first to give an exactly analytical expression for the energy current in hybrid nonlinear quantum structures.

\begin{acknowledgements}
This material is based upon work supported by the
National Science Foundation of China (Grants No. 11475160,
No. 61575180, No. 11575071), the Natural Science Foundation
of Shandong Province (No. ZR2014AM023, No.
ZR2014AQ026), the Basque Country Government (Grant No. IT986-16) and PGC2018-101355- B-I00 (MCIU/AEI/FEDER,UE).

\end{acknowledgements}

\end{document}